\documentclass{elsart1p}

\usepackage{epsfig}
\usepackage{subfigure}

\usepackage{amssymb}

\newcommand{\be}{\begin{equation}}
\newcommand{\ee}{\end{equation}}
\newcommand{\bea}{\begin{eqnarray}}
\newcommand{\eea}{\end{eqnarray}}

\begin{document}

\begin{frontmatter}

\vspace*{-1.0cm}




\title{Real-time gauge theory simulations from stochastic quantization with optimized updating}


\author{D\'enes Sexty}

\address{Institute for Nuclear Physics, Darmstadt University of Technology, Schlossgartenstr. 9, 64285 Darmstadt, Germany}

\begin{abstract}
Stochastic quantisation is applied to the problem of calculating 
real-time evolution on a Min\-kow\-ski\-an space-time lattice. 
We employ optimized updating using reweighting, or gauge fixing, respectively.
These procedures do not affect the underlying theory, but strongly improve
the stability properties of the stochastic dynamics. 
\end{abstract}

\begin{keyword}

\PACS 11.10.Wx \sep 04.60.Nc \sep 05.70.Ln \sep 11.15.Ha
\end{keyword}
\end{frontmatter}

\section{Introduction}
\label{sect:intro}

First-principles simulation of quantum field theories (such as QCD)
is a notoriously hard problem of theoretical physics.
Lattice calculations tipically use euclidean formulation,
where one can apply importance sampling.
In contrast, in Minkowski space, this is ineffective since the 
probability weight is in this case a phase factor: $ \exp ( i S ). $ 
Stochastic quantisation \cite{Parisi:1980ys}, however can be 
generalised to complex weights. Recently stochastic quantisation
was applied to calculate the real-time evolution of quantum fields
 (\cite{Berges:2005yt,Berges:2006xc} and references therein).
Recently stochastic quantisation has also been applied to nonzero 
chemical potential problems \cite{Aarts:2008rr}.
  In \cite{Berges:2007nr}, optimized updatings were studied, which 
improve the behaviour of the complex Langevin equations, whitout 
changing the underlying theory. The main insight is the usage of the 
fixedpoints of the Langevin flow as a criteria for convergence.

\section{Stochastic Quantisation}

The time evolution can be formulated using the path integral formalism
as an average weighted with $\exp(iS)$:

\be
\langle O(\Phi) \rangle = \int D\Phi e^{iS} O(\Phi) 
\ee

Using stochastic quantization the real-time quantum configurations in
3+1 dimensions are constructed by a stochastic process in an
additional (5th) Langevin-time $\theta$, by use of a Langevin equation:

\be \label{eq:complexlang}
{ \partial \Phi(\theta) \over \partial \theta } = 
 - i { \delta  S \over \delta \Phi } + \eta(\theta)
\ee

where $\eta$, the noise term satisfies:
$\langle \eta (\theta) \rangle = 0 , \   \langle \eta(\theta) \eta 
(\theta') \rangle = 2 \delta ( \theta - \theta')
$.
The expectation value of any observable can be calculated from 
the Langevin-time evolution
using the following formula:

\be 
\langle O(\Phi) \rangle = {1\over T} \int_0^T O(\Phi(\theta)) d \theta 
\ee

We applied stochastic quantisation to ``one plaquette'' models,
which serve to illustrate and test concepts of optimized updating,
a scalar oscillator (0+1 dimensional
scalar field theory), pure $SU(2)$ gauge theory, using 
the Wilson action on a 3+1 dimensional lattice.
The field theories were discretised on a comlex time contour, which enabled 
us to either calculate equilibrium distributions, or non-equilibrium 
time evolution. The contour tipically has a downward slope,
which improves convergence.

The complexity of the drift term in (\ref{eq:complexlang}) means 
that the originally real fields are complexified. This transfroms
a real scalar field to complex scalar, the link variables
of an $SU(N)$ field theory to $SL(N,{\bf C})$.
 Only after taking noise or Langevin-time averages,
respectively, the expectation values of the original real scalar theory 
($SU(N)$ gauge theory) are to be recovered. 
Accordingly, if the Langevin flow
converges to a fixed point solution of Eq.~(\ref{eq:complexlang})
it automatically fulfills the infinite hierarchy of
Dyson-Schwinger identities of the original
theory.

\section{Optimized updating of toy models}

As an example to illustrate reweighting we consider the one-plaquette 
model with $U(1)$
symmetry. For $U= e^{i \varphi}$ the action is given by
\begin{eqnarray}
S_0 \,=\,  \frac{\beta}{2} \left( U + U^{-1} \right) \,=\, \beta
\cos \varphi \,  \label{eq:ac0}
\end{eqnarray}
with real coupling parameter $\beta$. 
This simple model enables
us to check the results of the complex Langevin equation 
by evaluating the integral

\begin{eqnarray}
\langle O \rangle_0 \,=\, \frac{1}{Z_0} \int_{0}^{2\pi}{\rm
d}\varphi \, \e^{i S_0}\, O(\varphi) \, .\label{eq:av0}
\end{eqnarray}

For real $\beta$ the integrand in Eq.\ (\ref{eq:av0}) is not
positive definite, which mimics certain aspects of more
complicated theories in Minkowskian space-time. 
Solving the discretised Langevin equation corresponding to this model,
using $\beta=1$, and a Langevin stepsize $\epsilon=10^{-5}$, one gets:

\begin{equation}
\langle e^{i \varphi} \rangle_0 \, \stackrel{\rm without \atop
optimization}{=}\, -0.009(\pm 0.006) - i\, 0.00006(\pm 0.00007) \,
.
\end{equation}

Evaluating the numerical integral in (\ref{eq:av0}), one gets
$ \langle e^{i\varphi} \rangle \simeq i 0.575 $.  One observes that the simulation yields a wrong
result that is compatible with zero, in contrast to the
non-vanishing imaginary value obtained analytically.

The same averages may be calculated using by performing reweighting:
changing the action, and recompensating the change in the measurable so 
that the end-result remains the same.

\begin{eqnarray} \label{eq:rewe}
\langle O \rangle_0 \,=\, \frac{ \int_{0}^{2\pi}{\rm d}\varphi \,
\e^{i S_\alpha}\, e^{i(S_0 - S_\alpha) } \, O(\varphi)}{\int_{0}^{2\pi}{\rm
d}\varphi \, \e^{iS_\alpha} e^{i(S_0 - S_\alpha)   }}
\end{eqnarray}

\begin{figure}[h]
\begin{center}
\epsfig{file=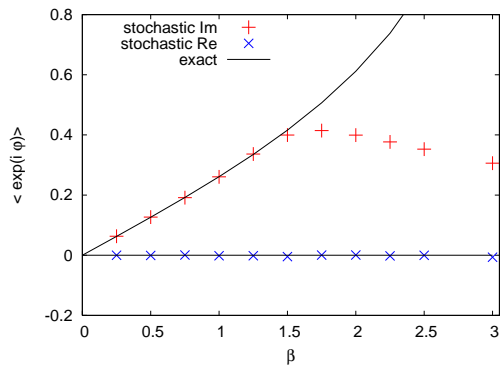,width=7.cm}
\end{center}
\vspace*{-0.0cm} \caption{\small The real and imaginary part of
the average $\langle e^{i \varphi} \rangle_{\alpha=1}$ as a
function of $\beta$. The lines represent averages obtained from
direct integration, while the symbols are measurements using a
stochastic process.} \label{fig:betadep}
\end{figure}

Evaluating the denominator and numerator of this factor one considers 
$S_\alpha$ as the action instead of $S_0$, thus the Langevin process is 
changed \cite{Berges:2007nr}. We will consider here the family of 
actions $S_\alpha=S_0 + \alpha \varphi=\beta \cos \varphi + \alpha \varphi$.
One finds that this new class of actions gives exact results 
for some region of its parameters (see Fig. \ref{fig:betadep}),
 such that  $\langle e^{i \varphi} \rangle_0$ is recoverable, 
using (\ref{eq:rewe}).

\begin{figure}[h]
\begin{center}
\epsfig{file=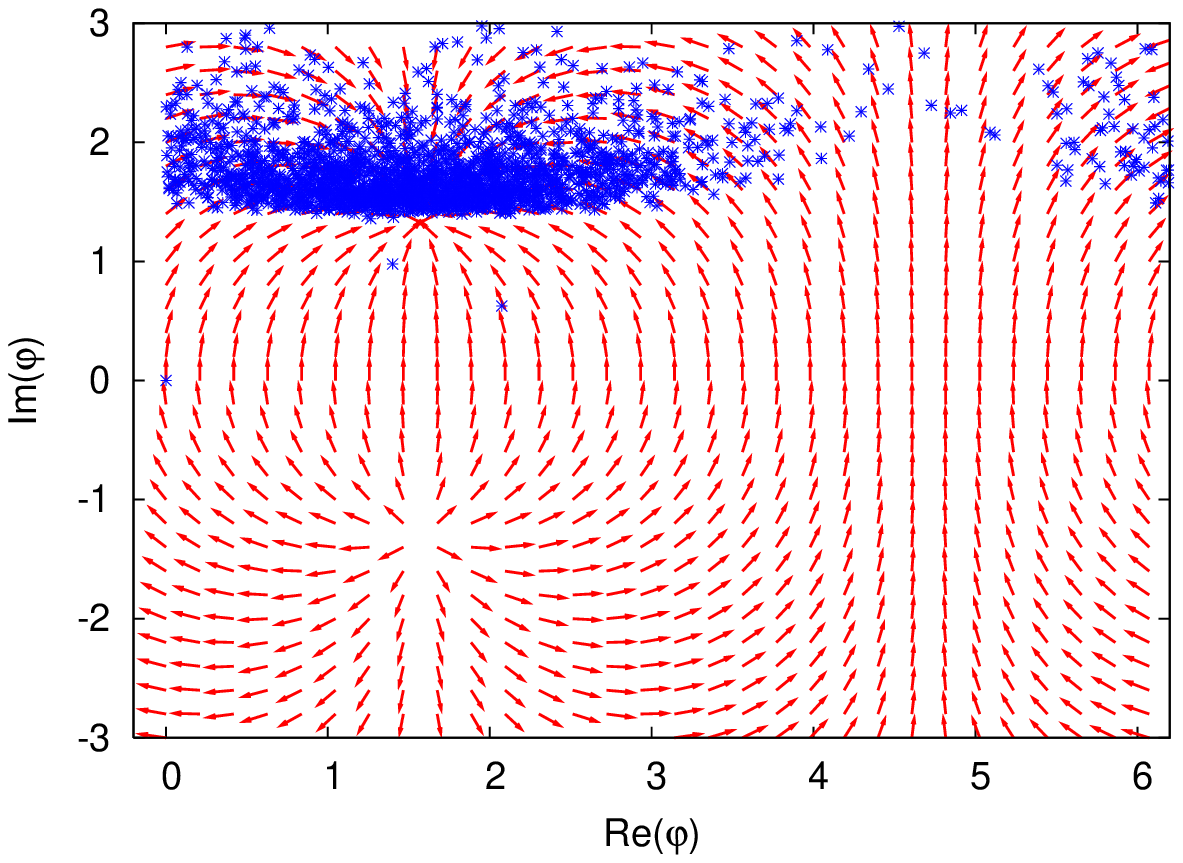,width=6.cm}
\epsfig{file=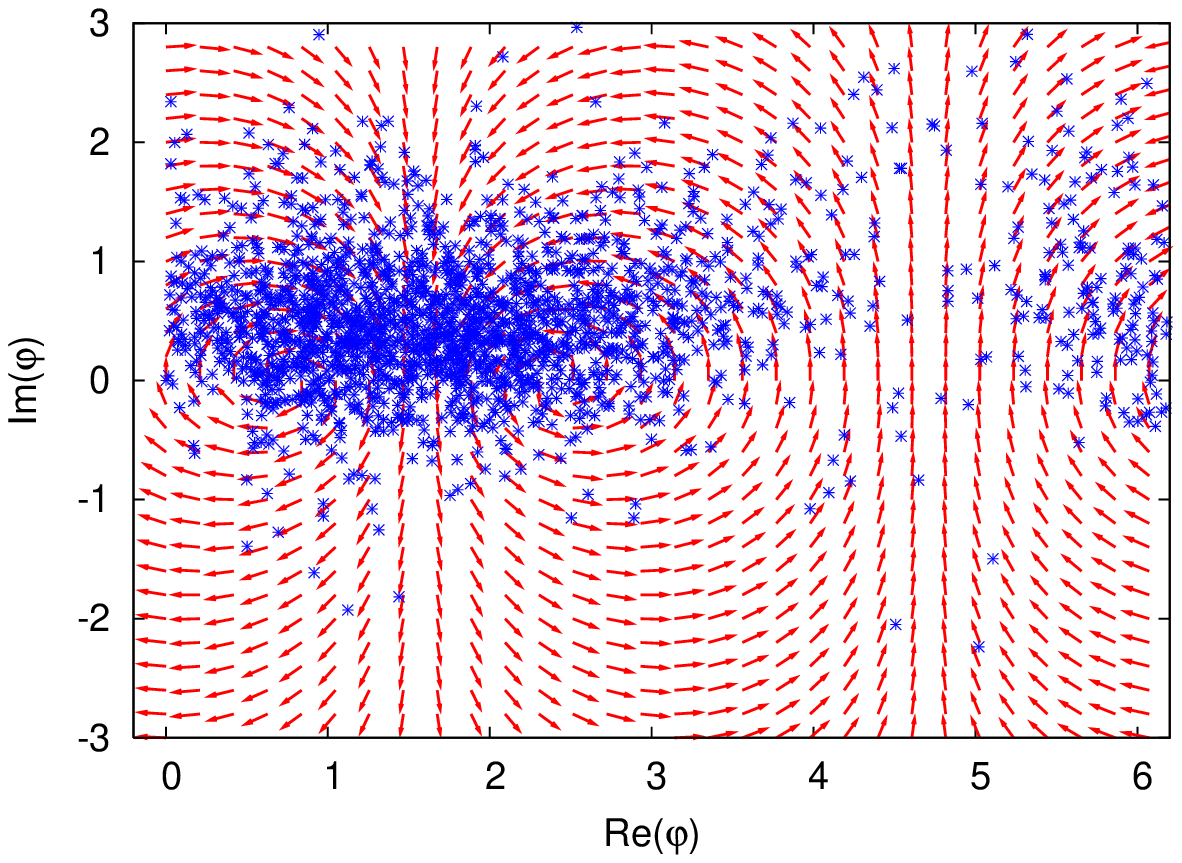,width=6.cm}
\end{center}
\vspace*{-0.1cm} \caption{\small  
Shown is the real and imaginary
part of $\partial S_\alpha/\partial \varphi$ plotted as a vector with
origin at each $\varphi$-value and with normalized length for
better visibility, using $\alpha=1$, and $\beta=0.5$ on the left panel 
and  $\beta=1.5$ on the right panel.
Also shown is the distribution of $\varphi$ as obtained from the full
solution of the respective complex Langevin equation. 
} 
\label{fig:fixedpoints}
\end{figure}

Qualitative understanding of the behavior of stochastic processes
corresponding to $S_\alpha$ is possible by studying the fixed point
structure of the drift term of the Langevin equation.  In
Fig. \ref{fig:fixedpoints}. the flowchart (normalized drift vectors)
is plotted alongside a scatter plot of $\varphi$ on the complex plane.
One observes (comparing with Fig. \ref{fig:betadep}) that the
stochastic results brake down when the fixed point structure
changes. Exact results correspond to an attractive fixedpoint, and a
relatively compact distribution of field values, while the absence of
attractive fixedpoints, and a wide distribution corresponds to the
breakdown. The breakdown of the process using the original $S_0$ 
action is consistent with
this picture: it has no attractive fixedpoints.

Consider now the $SU(2)$ one plaquette 
model, as a second example. The action is given by: 

\begin{eqnarray}
S(U) = \frac{\beta}{2}\, {\rm Tr}\, U \, ,\label{eq:ac2onepl}
\end{eqnarray}

which is invariant under the ``gauge'' transformation
$ U \rightarrow W^{-1} U W \label{eq:su2sym}$ with $U,W \in SL(2,{\bf C})$ 
 (after complexification).

\begin{figure}[t]
\begin{center}
\epsfig{file=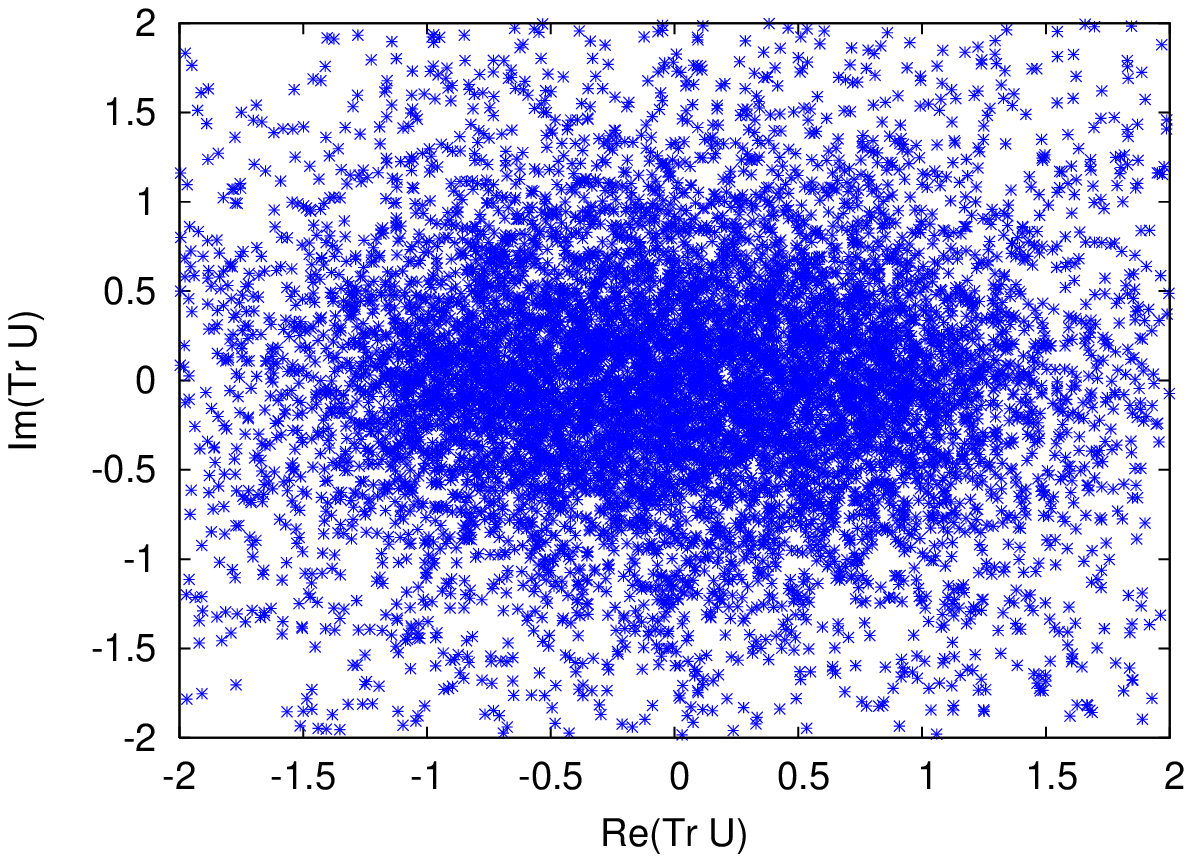,width=6.cm}
\epsfig{file=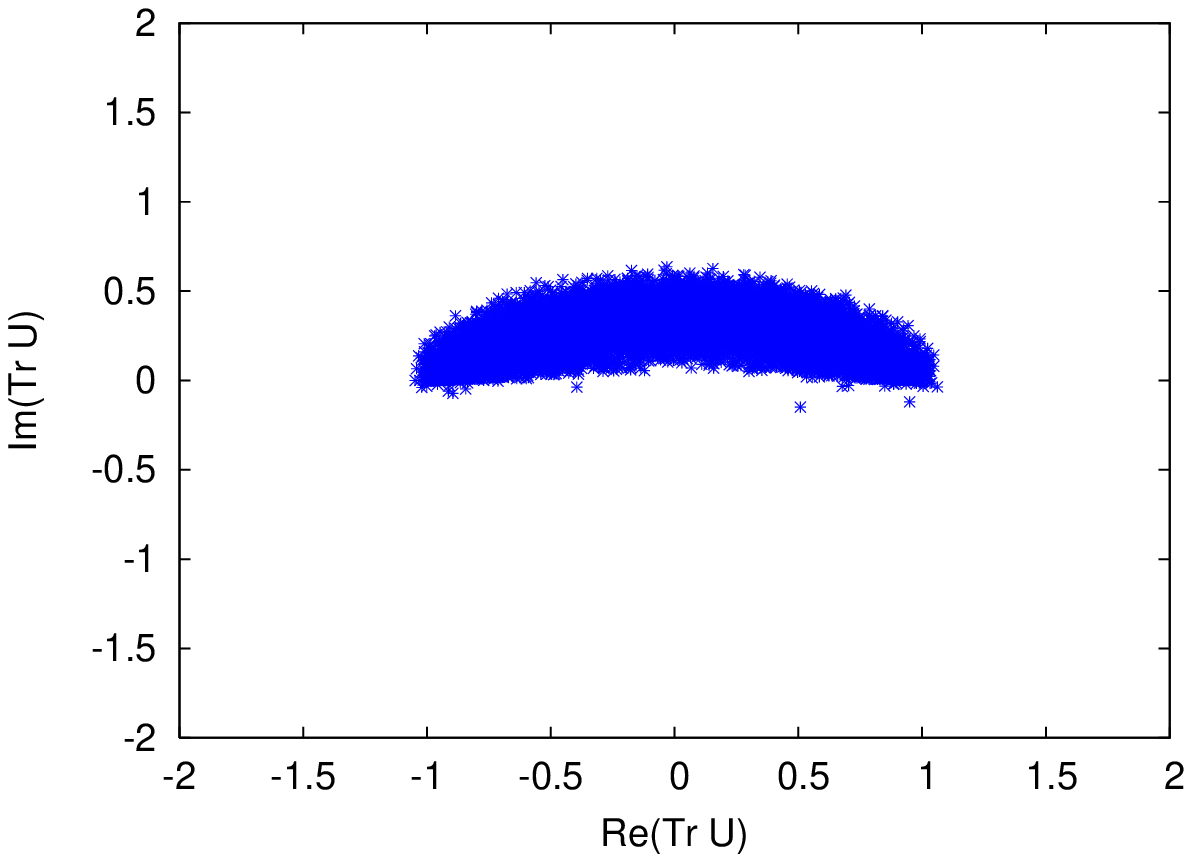,width=6.cm}
\end{center}
\vspace*{-0.5cm} \caption{\small The real and imaginary part of
the plaquette variable ${\rm Tr} U$ from snapshots with constant
Langevin-time stepping for the $SU(2)$ one-plaquette model with
$\beta = 1$. The left graph shows the wide distribution of values
obtained from the standard Langevin dynamics, while the right
graph displays the compact distribution from the ("gauge-fixed")
optimized process (see text for explanation).} \label{fig:distsu2}
\end{figure}

 Since the $SU(2)$ one-plaquette model has a
global symmetry, one may use this symmetry to "gauge-fix" certain
variables in order to constrain the growth of fluctuations. 
In the following we will use it in order to
diagonalize $U$ after each successive Langevin-time step: 
 $U^{\prime} \,=\, {\rm diag}\left(a+ i
\sqrt{1-a^2}, a - i \sqrt{1-a^2}\right) $ (for details, see 
\cite{Berges:2007nr}).
 The right graph of
Fig.~\ref{fig:distsu2} shows that the "gauge-fixing" leads to a
compact distribution, and exact results, in contrast to the distribution from
the unmodified process displayed on the left of that figure, which 
gives wrong results.

In conclusion, we have demonstrated the usage of optimized
updating on specific examples, which change the Langevin process by
reweighting or using the symmetries of the theory, in a way that it
gives exact results for the original theory.



{\bf Acknowledgements: } I would like to thank J\"urgen Berges, Szabolcs
Bors\'anyi and Ion-Olimpiu Stamatescu for a fruitful collaboration on 
stochastic quantization.

\end{document}